\documentclass[twocolumn,aps,preprintnumbers,amsmath,amssymb,superscriptaddress,nobibnotes]{revtex4}
\usepackage{graphicx}
\usepackage{dcolumn}
\usepackage{bm}
\usepackage{graphics}
\usepackage{longtable}
\usepackage{units}
\usepackage{textcomp}
\usepackage{float}

\begin{document}

\title{Correction of dephasing oscillations in matter wave interferometry}

\author{A. Rembold}
\affiliation{Institute of Physics and Center for Collective Quantum Phenomena in LISA$^+$,
University of T\"{u}bingen, Auf der Morgenstelle 15, 72076 T\"{u}bingen, Germany}
\author{G. Sch\"{u}tz}
\affiliation{Institute of Physics and Center for Collective Quantum Phenomena in LISA$^+$,
University of T\"{u}bingen, Auf der Morgenstelle 15, 72076 T\"{u}bingen, Germany}
\author{W.T. Chang}
\affiliation{Institute of Physics, Academia Sinica, Nankang, Taipei 11529, Taiwan, Republic of China}
\author{A. Stefanov}
\affiliation{Institute of Applied Physics, University of Bern, 3012 Bern, Switzerland}
\author{A. Pooch}
\affiliation{Institute of Physics and Center for Collective Quantum Phenomena in LISA$^+$,
University of T\"{u}bingen, Auf der Morgenstelle 15, 72076 T\"{u}bingen, Germany}
\author{I.S. Hwang}
\affiliation{Institute of Physics, Academia Sinica, Nankang, Taipei 11529, Taiwan, Republic of China}
\author{A. G\"{u}nther}
\email{a.guenther@uni-tuebingen.de}
\affiliation{Institute of Physics and Center for Collective Quantum Phenomena in LISA$^+$,
University of T\"{u}bingen, Auf der Morgenstelle 14, 72076 T\"{u}bingen, Germany}
\author{A. Stibor}
\email{alexander.stibor@uni-tuebingen.de}
\affiliation{Institute of Physics and Center for Collective Quantum Phenomena in LISA$^+$,
University of T\"{u}bingen, Auf der Morgenstelle 15, 72076 T\"{u}bingen, Germany}

\begin{abstract}
Vibrations, electromagnetic oscillations and temperature drifts are among the main reasons for dephasing in matter-wave interferometry. Sophisticated interferometry experiments, e.g. with ions or heavy molecules, often require integration times of several minutes due to the low source intensity or the high velocity selection. Here we present a scheme to suppress the influence of such dephasing mechanisms - especially in the low-frequency regime - by analyzing temporal and spatial particle correlations available in modern detectors. Such correlations can reveal interference properties that would otherwise be washed out due to dephasing by external oscillating signals. The method is shown experimentally in a biprism electron interferometer where a perturbing oscillation is artificially introduced by a periodically varying magnetic field. We provide a full theoretical description of the particle correlations where the perturbing frequency and amplitude can be revealed from the disturbed interferogram. The original spatial fringe pattern without the perturbation can thereby be restored. The technique can be applied to lower the general noise requirements in matter-wave interferometers. It allows for the optimization of electromagnetic shielding and decreases the efforts for vibrational or temperature stabilization.
\end{abstract}

\maketitle

\section{Introduction}

Matter-wave interferometers with electrons \cite{Mollenstedt1956a,Kiesel2002,Hasselbach1988a,Hasselbach2010}, atoms \cite{Carnal1991a,Keith1991a}, neutrons \cite{Rauch1974,Colella1975}, molecules \cite{Brezger2002a,Grisenti2000a} or ions \cite{Hasselbach1998a,Maier1997,Hasselbach2010} are all extremely sensitive to dephasing mechanisms. Thereby the phase of each single-particle wave is shifted relative to the detector by a temporal varying process. Integrating the individual interference patterns with such alternating phase shifts leads to a loss of contrast, even if full coherence is still maintained in the system. However, if the time-dependent phase shift were known, the dephasing could be corrected and full contrast could be recovered. Decoherence, on the other hand, causes a loss of contrast on the single-particle level, which cannot be corrected for.

The origin of dephasing mechanisms can be quite different, such as mechanical vibrations \cite{Stibor2005}, temperature drifts, or, especially important in the case of charged-particle interferometers, electromagnetic oscillations. While high frequency perturbations can be efficiently suppressed via vibrational isolation systems, electric filters, and mu-metal shieldings, low-frequency components become dominant. They can only be partially addressed by e.g using complex shielding schemes, low noise beam guiding electronics, and filtering of the \unit[50]{Hz} oscillation of the electric network. Moreover, the beam emission center might drift in position, as has been observed for conventional field ionization sources such as "supertips" \cite{Kalbitzer}, which was a major obstacle in the first realization of ion interferometry \cite{Hasselbach1998a,Maier1997,Hasselbach2010}. The suppression of low-frequency oscillations is therefore of major importance for the realization of stable particle interferometers. It is, for instance, a substantial challenge for the realization of sophisticated ion interference experiments \cite{Hasselbach1998a,Maier1997,Hasselbach2010}, e.g., in the context of Aharonov-Bohm physics \cite{Batelaan2009,Silverman1993}, where long signal integration times are necessary.

\begin{figure*}
\includegraphics[width=5.2in]{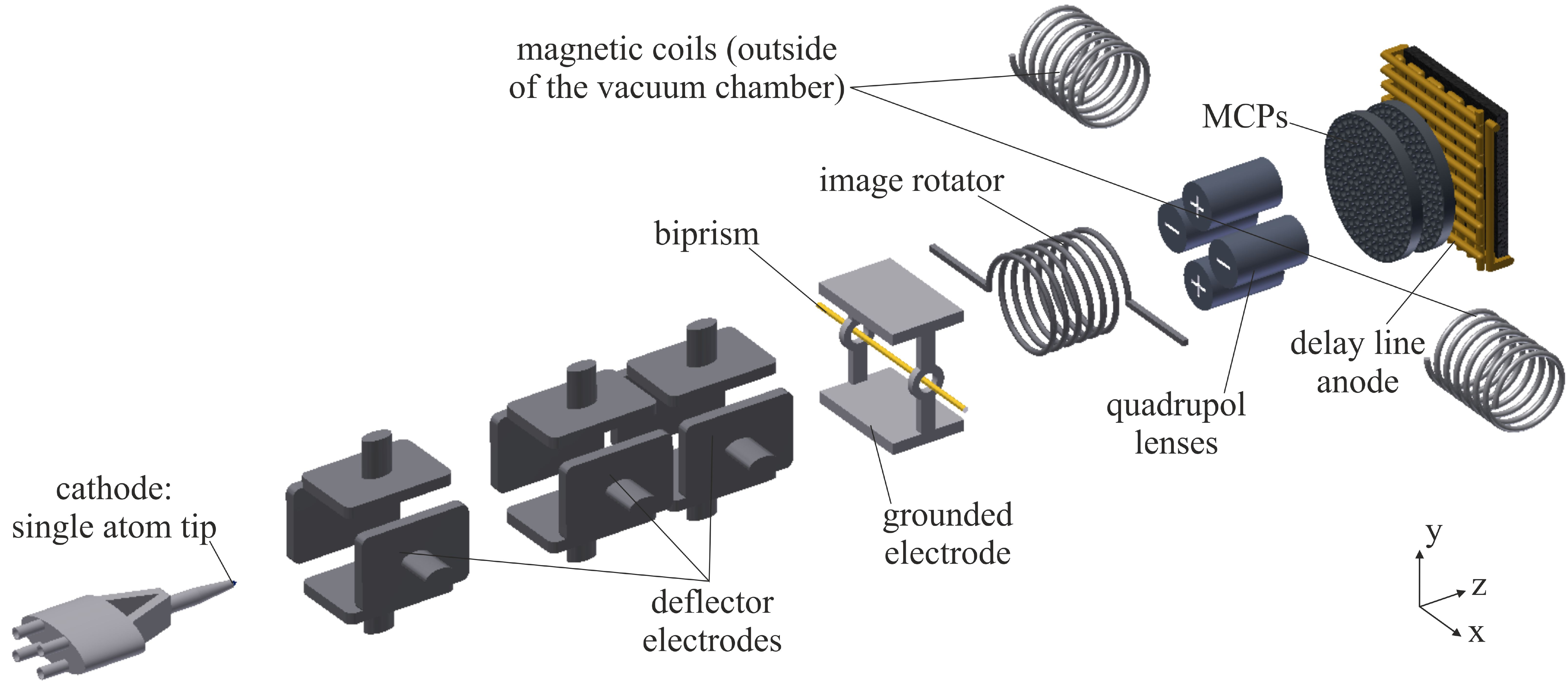} \caption{(Color)
In-vacuum setup of the electron biprism interferometer, which is a modified version of the one described in \cite{Hasselbach1998a,Maier1997,Hasselbach2010} (not to scale). An electron beam is field emitted from a single-atom tip \cite{Kuo2006a,Kuo2008} and adjusted by deflector electrodes towards the optical axis. The electron matter-waves are coherently separated and recombined by an electrostatic biprism \cite{Mollenstedt1956a}. The resulting interference pattern is magnified by quadrupole lenses and detected in a delay line detector \cite{Jagutzki2002}. It can be artificially disturbed by a time-varying field originating from two magnetic coils, which are placed outside the vacuum chamber. A mu-metal shield (not shown) is placed between the in-vacuum setup and the magnetic coils.}  \label{fig1}
\end{figure*}

In this article, we describe a method to significantly decrease the influence of low-frequency oscillations by including temporal and spatial particle correlations in the data analysis. The method is demonstrated experimentally using an electron interferometer, where a modern delay line detector \cite{Jagutzki2002} provides not only spatial information about the particle impact but also high temporal resolution. This makes them superior to commonly used multi-channel plates (MCPs) in conjunction with a fluorescence screen, which does not allow for high-precision time and position measurements. We show that, even after strong dephasing oscillations, the interference pattern can be recovered via correlation analysis. Therefore, we provide a full theory, which takes into account spatial and temporal correlations of all particle pairs. In principle, this method can be used for all periodic dephasing oscillations in the low-frequency regime below the particle count rate. Our method is thus of special importance in matter-wave experiments where temperature drifts or mechanical oscillations from the environment, such as the building, the cooling system, or the vacuum pumps, tend to wash out the interference pattern. For periodic perturbations our procedure can determine the frequency and amplitude of the dephasing signal and completely restore the spatial fringe pattern. The capability to identify the origin of dephasing is helpful for the design of further shielding or filtering in a matter-wave interferometer.

\section{Experiment}

We demonstrate the correlation analysis using a biprism electron interferometer. It was originally constructed by Hasselbach et al. \cite{Hasselbach1998a,Maier1997,Hasselbach2010} and modified with a new beam source and a new detector. Figure \ref{fig1} shows a sketch of the in-vacuum setup. It consists of an iridium covered tungsten (111) single-atom tip \cite{Kuo2006a,Kuo2008} that acts as a field emitter for a highly coherent electron beam. The field emission voltage was set to \unit[-1.53]{kV}. The vacuum pressure in the setup was \unit[$5\times10^{-9}$]{mbar}. The beam adjustment towards the optical axis of the setup is performed by using three deflector electrodes. Each one consists of four metal plates pairwise on opposite potentials to deflect the beam in the horizontal ($x$) and vertical ($y$) direction. The tip illuminates a fine gold coated biprism glass fiber that is oriented along $x$ and has a diameter of $\sim$ \unit[400]{nm} \cite{Warken2008}. It is positioned between two grounded plates to coherently split and recombine the electron matter-wave \cite{Mollenstedt1956a}. Setting the biprism on a positive potential of a few volts, the partial waves overlap and create an interference pattern parallel to the direction of the biprism fiber. For imaging purposes this interference pattern is expanded along $y$ using a quadrupole lens. A magnetic coil directly after the biprism is used as an image rotator to align the fringes to the quadrupole lens. The electron signal is amplified by a double-stage multichannel plate and detected by a delay line anode. The signal is recorded and analyzed by a computer. The whole in-vacuum setup is surrounded by a mu-metal shield, primarily damping high frequency electromagnetic perturbations.

Essential for our method of dephasing removal is the delay line detector. In biprism interferometry, interference patterns are typically detected using a MCP in conjunction with a fluorescent phosphor screen, which is then digitalized with a CCD camera. This allowed a moderate spatial resolution, restricted by the channel diameters of the MCP's, but only a rather limited temporal detection that is dependent on the long fluorescence decay time of the phosphor screen. With the delay line anode, single electrons can be detected with a spatial resolution below \unit[100]{\textmu m} and a time accuracy below \unit[1]{ns}. The dead time between two individual pulses is \unit[310]{ns} limiting the detectable count rate to the MHz regime. An incoming amplified electron pulse hits the delay line anode, consisting of two meandering wires oriented perpendicular to each other ($x$ and $y$ direction), and it induces a charge pulse in both directions of each wire. By measuring the time delay between those pulses at the wire endings, the spatial position and time of impact can be determined \cite{Jagutzki2002}.

\begin{figure*}
\includegraphics[width=1.0\textwidth]{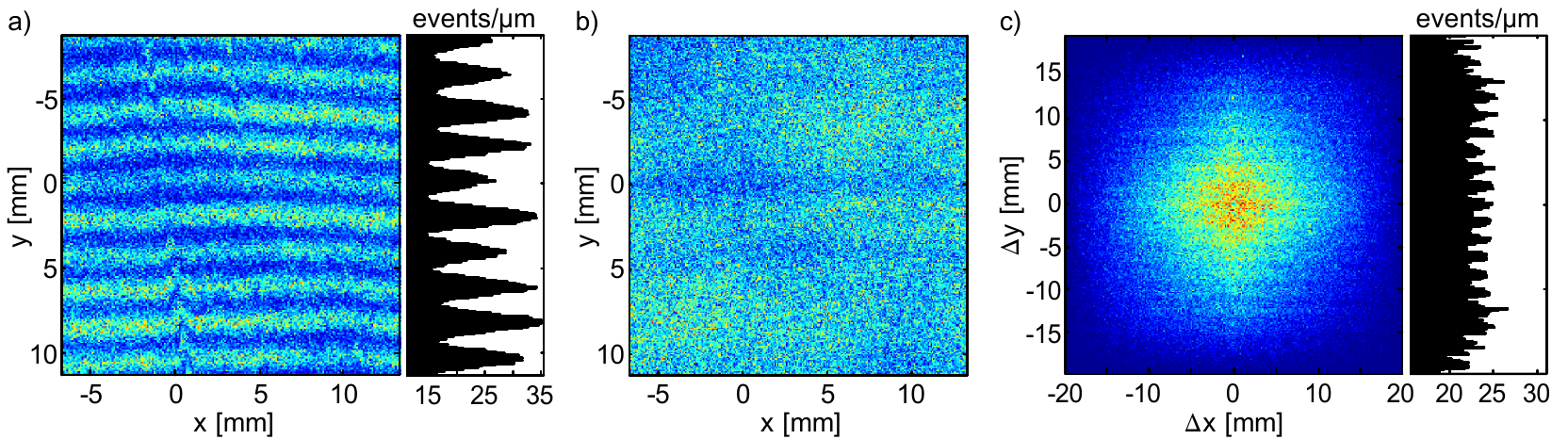} \caption{(Color) a) Electron biprism interference pattern and corresponding integrated line profile recorded with the setup of fig. \ref{fig1}b. The same interference pattern after the introduction of a periodic \unit[50]{Hz} magnetic field oscillation with a dc amplitude of 2$\pi$. The fringes are completely washed out. c) Histogram and integrated line profile of spatial distances $\Delta x$ and $\Delta y$ between temporal adjacent events. They clearly show correlations in the position of two consecutive events, revealing the existence of an interference pattern even after perturbation. The integrated line profile has been corrected by a factor $\left(1-|\Delta y|/Y\right)^{-1}$ to correct for the finite pattern width of \unit[$Y=20$]{mm}.}
\label{fig2}
\end{figure*}

Figure \ref{fig2}a shows an interference pattern, as detected with the delay line detector, after a total number of about $5\times 10^5$ electrons recorded in about \unit[100]{s}. This corresponds to a particle count rate of $\sim$ \unit[5]{kHz}. The distance between two fringes in the interferogram is $\sim$ \unit[2]{mm} and the contrast amounts to about \unit[35]{\%}. To demonstrate our method we artificially add a periodic dephasing in the form of an oscillating magnetic field. It is created by two external magnetic coils positioned outside the vacuum chamber and the mu-metal shield (see Figure \ref{fig1}). The magnetic field lines are oriented in the $x$-direction applying a force on the electrons in the $y$-direction normal to the interference fringes and parallel to the detection plane. This causes a periodic shift of the interference pattern, reducing the overall fringe contrast of the time integrated pattern. Using a function generator the frequency and amplitude of this disturbance can be tuned.

We disturbed the interference pattern of Figure \ref{fig2}a with a \unit[50]{Hz} oscillation. The amplitude was set such that it moved the pattern by \unit[$\pm$ 2]{mm} when a static current was applied to the coils. For oscillating currents, however, this amplitude is reduced due to the mu-metal shield around the in-vacuum setup of the interferometer. We thus expect a peak phase shift of the interferogram below $2\pi$. The resulting image with again $\sim 5\times 10^5$ events is illustrated in Figure \ref{fig2}b and clearly shows a washed out pattern where the contrast decreased to almost zero.

As our detector provides a list of coordinates and impact times of all consecutive incidents, we correlate each electron with its subsequent temporal neighbor by determining their spatial distance in the $x$- and $y$-direction ($\Delta x$, $\Delta y$). The relative commonness of these distances are plotted in Figure \ref{fig2}c. As it can be seen, the periodic pattern is revealed, a distinct evidence for matter-wave interferometry even in the presence of strong dephasing. For better visualization, Figure \ref{fig2}c includes the integrated line profile (corrected by the finite pattern width), which clearly shows a periodic modulation on the length scale of the original fringe pattern.

\section{Theory}

To gain more information about the disturbed interference pattern we apply a full correlation analysis on the data by including correlations not only between temporally adjacent particles, but also between all possible particle pairs. For the theoretical description of particle correlations in a periodically disturbed interference pattern, the probability distribution $f(y,t)$ of particle impacts at the detector is of major importance:

\begin{eqnarray}
f(y,t) & = & f_0\left(1+K\cos \left(ky + \varphi\left(t\right)\right)\right) \label{eq1} \\
\quad \mbox{with}\quad \varphi(t) & = & \varphi_0\cos(\omega t) \nonumber
\end{eqnarray}

\noindent At each time $t$, the distribution function is normalized via $f_0$ and characterized by its spatial periodicity $\lambda=2\pi/k$, contrast $K$, and phase $\varphi$. The time dependence of $\varphi$ causes a periodic phase shift of the probability distribution, with $\varphi_0$ being the peak phase deviation and $\omega=2\pi\nu$ the frequency of the phase disturbance. The corresponding interference pattern observed at the detector is given by the time averaged probability distribution
\begin{equation}
\lim_{T\rightarrow\infty} \frac{1}{T}\int_0^T f(y,t) dt \,=\, f_0\left(1+K\,J_0(\varphi_0)\,\cos(ky)\right),\label{eq2}
\end{equation}
with $J_0$ being the zero-order Bessel function of the first kind. Depending on the strength of the phase deviation, the visible contrast is thus reduced by a factor of $J_0(\varphi_0)$, leading to vanishing interference fringes for large $\varphi_0$.

\begin{figure*}
\includegraphics[width=1.0\textwidth]{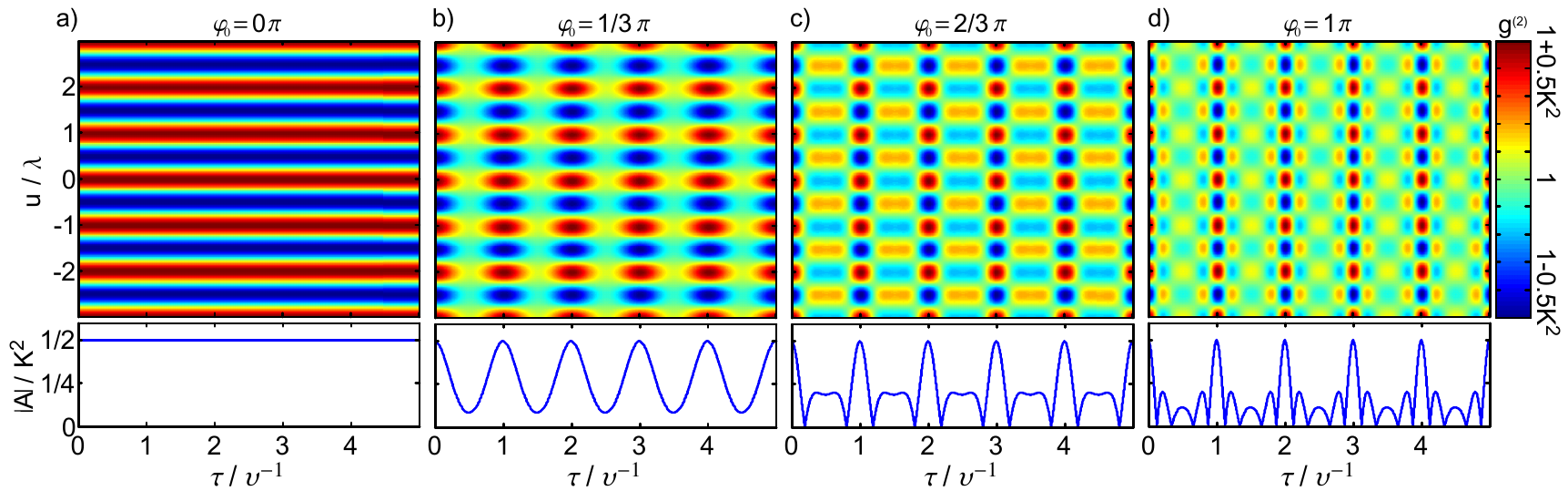} \caption{(Color online) Two-dimensional correlation functions $g^{(2)}(u,\tau)$ of disturbed interference patterns as a function of the spatial ($u$) and temporal ($\tau$) distance between two detection events. The axes are normalized to the spatial periodicity ($\lambda$) and the perturbation period ($\nu^{-1}$). The plots visualize the results of Eq. \ref{eq5} for different peak phase deviations $\varphi_0$ of (a) $0\pi$, (b) $1/3\pi$, (c) $2/3\pi$, and (d) $1\pi$. The absolute value of $A(\tau)$ normalized to the contrast $K^2$, is shown below each plot.}
\label{fig3}
\end{figure*}

If the detector allows for spatial and temporal information on the particle arrival, a correlation analysis can be used to retain information of the undisturbed interference pattern and the phase disturbance. Starting from eq. \ref{eq1}, the second order correlation function reads
\begin{equation}
g^{(2)}(u,\tau) = \frac{\ll f(y+u,t+\tau) f(y,t)\gg_{y,t}}{\ll f(y+u,t+\tau) \gg_{y,t} \ll f(y,t) \gg_{y,t}},
\label{eq3}
\end{equation}
where $\ll \cdot \gg_{y,t}$ denotes the average over position and time
\begin{equation}
\ll f(y,t) \gg_{y,t} = \lim_{Y,T\rightarrow\infty} \frac{1}{TY}\int_0^T \int_{-Y/2}^{Y/2} f(y,t) \,dy\, dt\label{eq4}
\end{equation}
In the limit of large acquisition times $T\gg 2\pi/\omega$ and lengths $Y\gg \lambda$ the integrals can be solved and the correlation function becomes
\begin{equation}
g^{(2)}(u,\tau) = 1 \,+\, A(\tau)\cos(ku) \label{eq5}
\end{equation}
with
\begin{eqnarray}
A(\tau) & = & \frac{1}{2}K^2\sum_{n=-\infty}^{\infty} J_n(\varphi_0)^2 \, \mbox{e}^{-in\omega\tau}\label{eq6}\\
& = & \frac{1}{2}K^2 J_0(\varphi_0)^2 \, + \, K^2\sum_{n=1}^{\infty}J_n(\varphi_0)^2\cos(n\omega\tau)\label{eq7}
\end{eqnarray}

Centered around 1, the second order correlation function of the disturbed interference pattern thus shows a periodic modulation in the spatial distance $u$ between two detection events with the same  periodicity as the undisturbed interference pattern. The amplitude of this modulation, however, depends on the correlation time $\tau$. In the frequency domain $A(\tau)$ can be decomposed to a superposition of sidebands at discrete frequencies $n\omega$ ($n\in \mathbb{Z}$) with strengths given by the peak phase deviation and the Bessel functions $J_n(\varphi_0)$.

Figure \ref{fig3} shows the two-dimensional correlation function and the amplitude $|A(\tau)|$ for different peak phase deviations. As illustrated in Figure \ref{fig3}a, without modulation (undisturbed interference pattern) only $J_0$ remains non-zero and $A=0.5\,K^2$. The correlation function thus becomes independent of $\tau$ and resembles the original interference pattern (Figure \ref{fig2}a). For small but non-zero $\varphi_0$, the first order Bessel function $J_1$ comes into play causing a sinusoidal modulation of $A$ at frequency $\omega$ (Figure \ref{fig3}b). As $\varphi_0$ increases further, more and more higher-order Bessel functions have to be taken into account, adding higher harmonic modulations to $A$. However, maximal spatial contrast $0.5 K^2$ is only recovered at multiples of $\tau=1/\nu$, where all higher harmonics constructively interfere (Figure \ref{fig3}c and d). Independent of the peak phase deviation, the correlation analysis thus reveals the frequency of the phase disturbance and the spatial frequency of the underlying interference pattern.

Before the correlation theory can be applied on our measurements, the second order correlation function needs to be extracted from the detector signal. This signal is given by the position $y_i$ and time $t_i$ of all particle events $i=1\ldots N$. Following eq. \ref{eq3} the correlation function is basically determined by the number $N_{u,\tau}$ of particle pairs $(i,j)$ with $\left(t_i-t_j\right) \in \left[\tau,\tau+\Delta \tau\right]$ and $\left(y_i-y_j\right) \in \left[u,u+\Delta u\right]$
\begin{equation}
g^{(2)}(u,\tau) = \frac{TY}{N^2 \Delta \tau \Delta u}\,\frac{N_{u,\tau}}{\left(1-\frac{\tau}{T}\right)\left(1-\frac{\left|u\right|}{Y}\right)}
\label{eq8}
\end{equation}
with normalization factor $TY/N^2$ and discretisation step size $\Delta \tau$ and $\Delta u$. The additional factor $\left[(1-\tau T^{-1})(1-\left|u\right|Y^{-1})\right]^{-1}$ corrects $N_{u,\tau}$ for the finite acquisition time $T$ and length $Y$ because large time and position differences will be less likely to be observed.

\begin{figure*}
\includegraphics[width=1.0\textwidth]{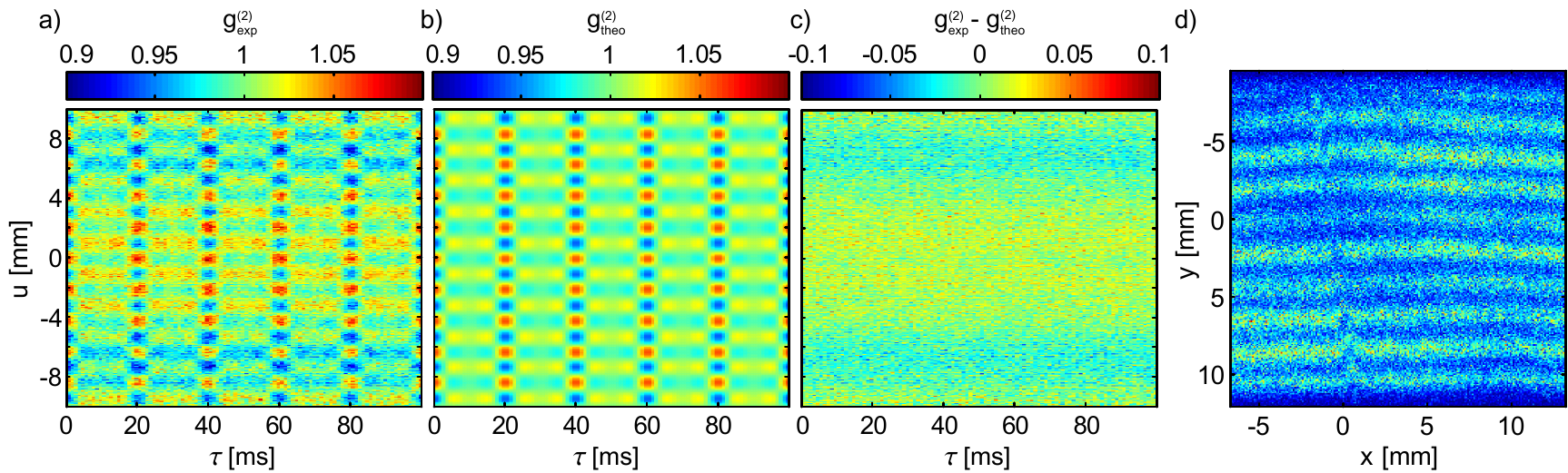} \caption{(Color online) a) Experimentally determined two dimensional correlation function $g^{(2)}(u,\tau)$, extracted according to Eq. \ref{eq8} from the detector signal of the disturbed interference pattern in Figure \ref{fig2} (b). (b) Fit of the pattern in (a) with the theoretical expression in Eqs. \ref{eq5}-\ref{eq7}. (c) The residual of theoretical and experimental data. The remaining periodic structure might be related to diffraction at the biprism. (d) Reconstruction of the original undisturbed interference pattern from the strongly disturbed data of \ref{fig2}b using the extracted fitting parameters.}
\label{fig4}
\end{figure*}

\section{Results}

To apply the theory to the outcome of our electron biprism experiment we extracted the $g^{(2)}(u,\tau)$ function according to Eq. \ref{eq8} from the raw data corresponding to Figure \ref{fig2}b. The result is shown in Figure \ref{fig4}a. As described in Sec. III, the periodicity of this pattern in $u$ and $\tau$ is an apparent sign of matter-wave interference that can be observed even in experimental conditions with significant periodic dephasing perturbations.

Using the theoretical expression of eq. \ref{eq5}-\ref{eq7}, we fitted the data in Figure \ref{fig4}a. The fit and the remaining residuum are shown in Figure \ref{fig4}b and \ref{fig4}c, respectively. They reveal all parameters describing the interferogram and the perturbation. The fitted parameters are: \mbox{$\nu=\;$\unit[49.996 ($\pm ~0.018$)]{Hz}} for the dephasing frequency, \mbox{\unit[$K=34.5$ ($\pm ~0.2$)]{\%}} for the interference contrast, \mbox{\unit[$\lambda=2.089$ ($\pm ~0.001$)]{mm}} for the spatial period of the interference pattern, and \mbox{\unit[$\varphi_0=0.802$ ($\pm ~0.004$)]{$\pi$}} for the peak phase deviation. These values are in excellent agreement with the properties of the unperturbed interference pattern (\unit[$K\approx 35$]{\%}, \unit[$\lambda\approx 2$]{mm}) and the applied disturbance frequency (\unit[$\nu=50$]{Hz}). Only the peak phase deviation shows a discrepancy to its dc value of $\varphi_0=2\pi$, which is due to the mu-metal shield between the interferometer and the dephasing coils. As expected, this shield damps the amplitude of external field oscillations.

The lack of any sub-structure in the residual plot (Figure \ref{fig4}c) shows that our correlation model is well suited to describe the experimental data. The residuum shows only a weak remaining structure on the length scale of $\sim$ \unit[7]{mm}, which is probably due to diffraction on both edges of the biprism.

We demonstrated that it is possible to extract the unknown frequency and amplitude of periodic, single frequency dephasing oscillations from the perturbed interference pattern even if no interference fringes are visible in the spatially integrated image (Figure \ref{fig2}b). With the obtained parameters and the spatial and temporal coordinates of the events, we are able to reverse the perturbation. This can be done by shifting each event back to its original, undisturbed coordinate according to the determined information,
\begin{equation}
y_{new} = y \, - \, \frac{\lambda}{2\pi}\varphi_0 \cos(\omega t + \phi) ~.
\label{ynew}
\end{equation}
The only parameter we do not obtain from the fit is the starting phase of the perturbation $\phi$. We extract it by varying the starting phase between 0 and $2\pi$ until the maximal contrast of the resulting interference pattern is achieved. Figure \ref{fig4}d shows the reconstructed interference pattern. It agrees well with the experimentally undisturbed pattern of Figure \ref{fig2}a. Even small structures like the local phase shifts by charged dust particles on the biprism can be reconstructed.

\newpage

\section{Conclusion}

Sensitive and accurate matter-wave interference experiments are susceptible to dephasing perturbations that wash out the interference pattern and decrease the contrast \cite{Stibor2005}. The dephasing can be due to electromagnetic oscillations, electrical network oscillations, temperature drifts or mechanical vibrations. Usually major efforts to shield or damp these setups are required to suppress these perturbations.

We have presented a method to effectively decrease dephasing effects by including temporal and spatial correlations between the detected particles in the analysis of an interference signal. The full correlation analysis reveals the fringe pattern even in the presence of oscillating perturbations, while conventional methods that rely only on spatial signal accumulation are not able to verify matter-wave interference. The analysis can be applied whenever the frequency of the perturbing signal is significantly lower than the average incident rate on the detector. This condition is well met for most interference experiments since signal rates of several kHz and perturbations below a few hundred Hz are common. Besides information on the perturbation, our method can be used to retain the undisturbed interference pattern.

Our method has potential applications in any kind of charged and neutral particle interferometer where a detector with a high spatial as well as temporal resolution is used. Nowadays, such detectors are available for electrons \cite{Jagutzki2002}, ions \cite{Jagutzki2002}, neutrons \cite{Siegmund2007} and neutral atoms \cite{Schellekens2005}. The technique  is of general importance for the analysis of dephasing perturbations in matter-wave interferometry, and it allows to optimize shielding and damping installations. It decreases the requirements for vibrational stabilization, temperature stabilization and filtering of low-frequency perturbations from electronic instruments.

\section{Acknowledgements}

This work was supported by the Deutsche Forschungsgemeinschaft (DFG, German Research Foundation) through the Emmy Noether program STI 615/1-1.A.R. acknowledges support from the Evangelisches Studienwerk e.V. Villigst, W.T.C. and I.S.H. from the Academia Sinica project AS-102-TP-A01, A.Ste. acknowledges the support from the Swiss National Science Foundation and A.G. from the DFG SFB TRR21. The authors thank P. Federsel, H. Prochel, and F. Hasselbach for helpful discussions.

\end{document}